\documentclass[sigconf, nonacm]{acmart}
\setcopyright{none}
\usepackage{multirow}
\usepackage{placeins}

\usepackage{booktabs}
\usepackage{colortbl}
\usepackage{xcolor}
\usepackage{pifont}   
\usepackage{array}
\usepackage{graphicx}
\definecolor{lightgray}{gray}{0.94}

\definecolor{lightgray}{gray}{0.94}

\AtBeginDocument{%
  }

\setcopyright{acmlicensed}
\copyrightyear{2025}
\acmYear{2025}
\acmDOI{XXXXXXX.XXXXXXX}
\acmConference[]{}
\acmISBN{}

\acmSubmissionID{}

\begin{document}

\title{PresentCoach: Dual-Agent Presentation Coaching through Exemplars and Interactive Feedback}

\author{Sirui Chen}
\authornote{Contributed equally to this research}
\affiliation{%
  \institution{The Hong Kong University of Science and Technology (Guangzhou)}
  \city{Guangzhou}
  \state{Guangdong}
  \country{China}}
\email{schen488@connect.hkust-gz.edu.cn}

\author{Jinsong Zhou}
\authornotemark[1]
\affiliation{%
  \institution{The Hong Kong University of Science and Technology (Guangzhou)}
  \city{Guangzhou}
  \state{Guangdong}
  \country{China}}
\email{jzhou945@connect.hkust-gz.edu.cn}

\author{Xinli Xu}
\authornotemark[1]
\affiliation{%
  \institution{The Hong Kong University of Science and Technology (Guangzhou)}
  \city{Guangzhou}
  \state{Guangdong}
  \country{China}}
\email{xxu068@connect.hkust-gz.edu.cn}

\author{Xiaoyu YANG}
\affiliation{%
  \institution{The Hong Kong University of Science and Technology (Guangzhou)}
  \city{Guangzhou}
  \state{Guangdong}
  \country{China}}
\email{xyang058@connect.hkust-gz.edu.cn}

\author{Litao Guo}
\affiliation{%
  \institution{The Hong Kong University of Science and Technology (Guangzhou)}
  \city{Guangzhou}
  \state{Guangdong}
  \country{China}}
\email{guolitauo@gmail.com}

\author{Ying-Cong Chen}
\authornote{Corresponding author}
\affiliation{%
  \institution{The Hong Kong University of Science and Technology (Guangzhou)}
  \city{Guangzhou}
  \state{Guangdong}
  \country{China}}
\email{yingcongchen@ust.hk}
\renewcommand{\shortauthors}{Chen et al.}

\begin{abstract}
Effective presentation skills are essential in education, professional communication, and public speaking, yet learners often lack access to high-quality exemplars or personalized coaching. Existing AI tools typically provide isolated functionalities—such as speech scoring or script generation—without integrating reference modeling and interactive feedback into a cohesive learning experience. We introduce a dual-agent system that supports presentation practice through two complementary roles: the Ideal Presentation Agent and the Coach Agent. The Ideal Presentation Agent converts user-provided slides into model presentation videos by combining slide processing, visual-language analysis, narration script generation, personalized voice synthesis, and synchronized video assembly. The Coach Agent then evaluates user-recorded presentations against these exemplars, conducting multimodal speech analysis and delivering structured feedback in an Observation–Impact–Suggestion (OIS) format. To enhance the authenticity of the learning experience, the Coach Agent incorporates an Audience Agent, which simulates the perspective of a human listener and provides humanized feedback reflecting audience reactions and engagement. Together, these agents form a closed loop of observation, practice, and feedback. Implemented on a robust backend with multi-model integration, voice cloning, and error handling mechanisms, the system demonstrates how AI-driven agents can provide engaging, human-centered, and scalable support for presentation skill development in both educational and professional contexts.
\end{abstract}

\ccsdesc[500]{Do Not Use This Code~Generate the Correct Terms for Your Paper}
\ccsdesc[300]{Do Not Use This Code~Generate the Correct Terms for Your Paper}
\ccsdesc{Do Not Use This Code~Generate the Correct Terms for Your Paper}
\ccsdesc[100]{Do Not Use This Code~Generate the Correct Terms for Your Paper}

\keywords{AI coaching, presentation training, multimodal feedback, human–AI interaction}
\begin{teaserfigure}
  \includegraphics[width=\textwidth]{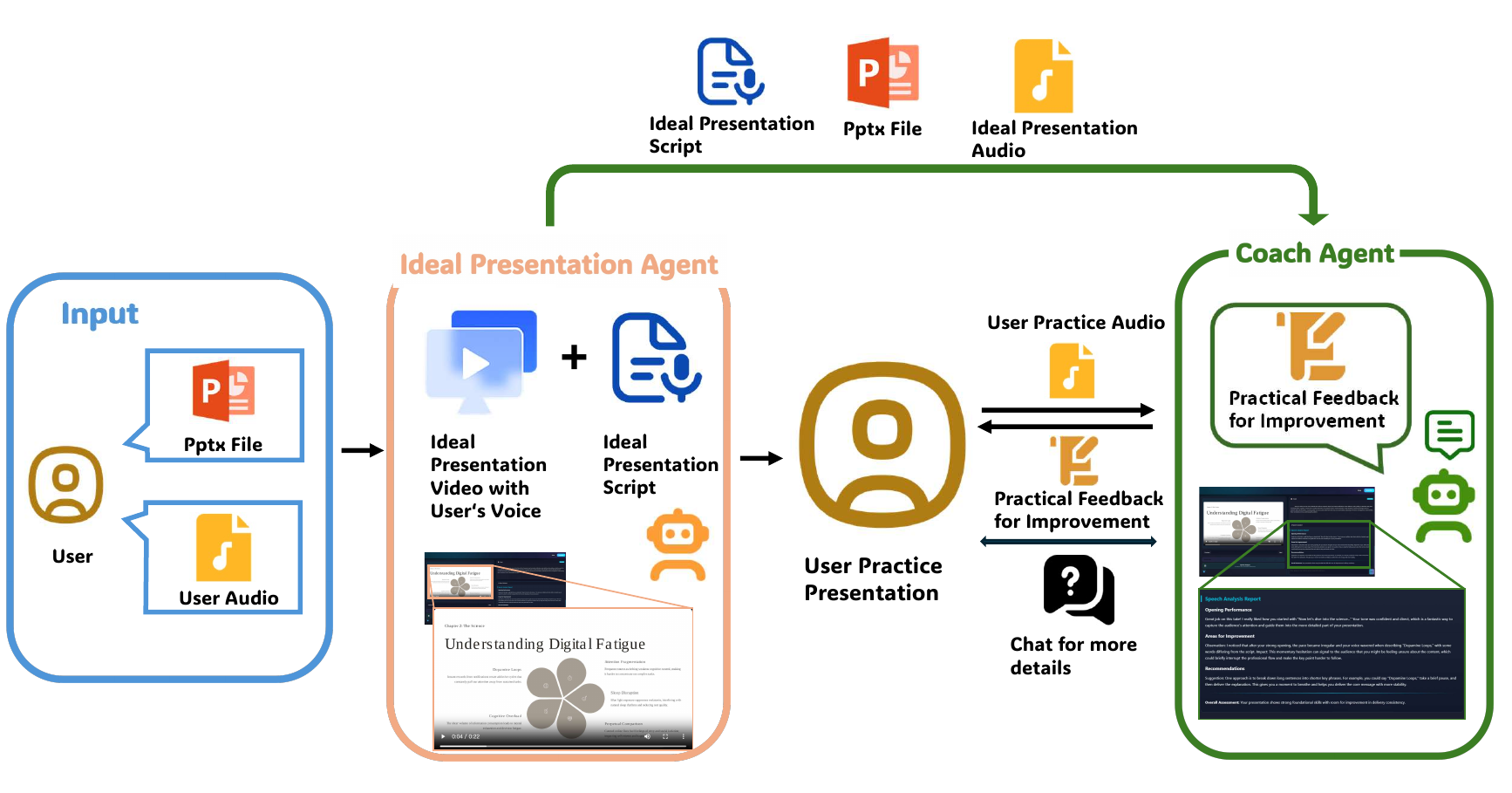}
  \caption{The Simple Workflow of our PresentCoach system. Initially, a user provides a .pptx file and an audio sample. The Ideal Presentation Agent processes these inputs to create a benchmark presentation, complete with an ideal script and narration in the user's voice. This serves as a standard for the user to practice their delivery. Subsequently, the user can submit their practice audio to the Coach Agent for evaluation and feedback. A chat feature is also available for in-depth discussion with the Coach Agent.}
  \label{fig:teaser}
\end{teaserfigure}

\received{}
\received[revised]{}
\received[accepted]{}

\maketitle
\section{Introduction}
Effective presentation skills are essential for education, communication, and career advancement\cite{vzivkovic2014importance, bradbury2006successful}.
As public speaking and digital presentations become integral to professional and academic settings, there is growing interest in intelligent systems that can assist users in improving delivery, confidence, and clarity.
Recent progress in speech analytics~\cite{chu2023qwen,radford2023robust}, virtual tutors~\cite{hu2024gaussianavatar,qian20243dgs}, and multimodal AI~\cite{achiam2023gpt,Qwen3-Omni,gemini25pro} has made it possible to analyze and even simulate human presentation behavior.
Despite these advances, effective presentation training remains difficult to scale~\cite{hoque2013mach}.
Traditional approaches such as workshops, peer feedback, or personalized coaching require substantial time, expertise, and cost~\cite{wortwein2015multimodal}.
Commercial tools such as Yoodli~\cite{yoodliWebsite} and Poised~\cite{poisedWebsite} provide automated feedback but primarily focus on surface-level metrics such as pacing, filler words, and prosody~\cite{fung2015roc,hoque2013mach}.
These systems lack the pedagogical depth to support deliberate practice, iterative reflection, and contextual understanding of slide content~\cite{chi2009active}.

Academic research has explored AI-based presentation training through various modalities.
Systems such as ROCSpeak~\cite{fung2015roc} and PresentationTrainer~\cite{schneider2015presentation} provide real-time prosody or gesture feedback,
while recent frameworks like PASS~\cite{aggarwal2025pass} and PresentAgent~\cite{shi2025presentagent} use large language models to generate exemplar scripts or videos.
These works represent significant progress in automating aspects of presentation support.
However, most existing systems treat presentation learning as a one-shot process that either generates exemplar content or scores delivery performance without integrating both demonstration and feedback~\cite{chinn2012model}.
As a result, learners lack an explicit reference for “what good looks like” and receive fragmented numerical metrics rather than contextual and actionable guidance.
In addition, current systems seldom simulate audience perception~\cite{bailenson2006transformed}, which is an essential factor for effective communication and confidence building.

To address this challenge, we draw inspiration from how effective speakers learn: study a strong exemplar, rehearse with intent, then refine through targeted feedback. Humans excel at learning from demonstration~\cite{gaskins2010learning} because clear exemplars render tacit standards visible, accelerate calibration of tone, pacing, and slide-speech alignment, and shorten the path to a first competent draft. Complementing this, targeted feedback concentrates the search for improvement by turning vague impressions into specific next steps, clarifying what to change, why it matters, and how to do it, so deliberate practice yields rapid, iterative gains.

In this work, we introduce \textit{PresentCoach}, a dual-agent system for presentation practice with two key innovations. First, the Ideal Presentation Agent produces a slide-aware exemplar video for the user’s own deck by interpreting visual content, drafting a concise narration, cloning a consistent voice, and assembling synchronized slide–audio segments. This makes “what good looks like” concrete and directly comparable. Second, the Coach Agent provides structured, conversational feedback using an Observation-\allowbreak Impact-\allowbreak Suggestion (OIS) scheme and an integrated Audience Agent that translates analytics into humanized listener reactions. Together, these components create a localized, iterative learning loop in which learners observe a tailored exemplar, rehearse targeted slides or segments, and refine delivery with actionable, audience-aware guidance. A comparison between PresentCoach and existing academic and commercial presentation coaching systems is summarized in Table~\ref{tab:comparison}.

To evaluate these ideas, we conducted a controlled between-subjects user study (\textit{N}=24) comparing AI-assisted rehearsal (\textbf{PresentCoach+PPT}) with self-directed slide practice (\textbf{PPT-only}). 
The study investigated both affective and experiential outcomes in a realistic, time-constrained rehearsal context.  
Our investigation was guided by three research questions:  

\begin{itemize}
    \item \textbf{RQ1:} Compared with \textit{PPT-only} self-practice, does rehearsal with \textit{PresentCoach+PPT} reduce public-speaking anxiety?  
    \item \textbf{RQ2:} While providing richer, feedback-intensive guidance, does \textit{PresentCoach+PPT} maintain an acceptable cognitive workload and achieve satisfactory post-use usability?  
    \item \textbf{RQ3:} How do participants perceive and experience the collaboration between the Ideal Presentation Agent and the Coach Agent during rehearsal?
\end{itemize}

Quantitative results show that \textit{PresentCoach} significantly increased self-reported speaking confidence, as measured by the PRCS scale, compared with the control group. 
NASA-TLX~\cite{hart1988development} ratings indicated that the system imposed only a moderate workload, while SUS~\cite{brooke1996sus} scores exceeded established usability benchmarks. 
Participants also rated the collaboration between the two agents positively, highlighting clear role differentiation, complementary guidance, and motivational balance. 
Qualitative analyses of post-session interviews further revealed that the integration of exemplar modeling and feedback coaching created a psychologically safe and motivating environment that encouraged deliberate practice and reflection.

In summary, our contributions of this paper are three-fold:
\begin{itemize}
    \item We present \textit{PresentCoach}, a dual-agent, end-to-end system for presentation training that operationalizes the \emph{observation $\rightarrow$ practice $\rightarrow$ reflection} cycle within a single interface.

    \item The system features two complementary agents: an \emph{Ideal Presentation Agent} that generates slide-aware exemplar videos in the user’s cloned voice, and a \emph{Coach Agent} that provides structured  feedback with an audience-aware mechanism simulating listener reactions.

    \item A controlled study shows that \textit{PresentCoach} reduces speaking anxiety, maintains moderate cognitive workload, and achieves high usability compared with self-practice. Participants perceived the two agents as distinct yet complementary—balancing motivation and guidance—and reported increased confidence and engagement through their interaction.
\end{itemize}



\section{Related Work}

\begin{table}[t]
\centering
\caption{\textbf{Comparison of PresentCoach with existing presentation coaching systems and commercial tools.}}
\renewcommand{\arraystretch}{1.1}
\setlength{\tabcolsep}{4pt}
\resizebox{0.9\linewidth}{!}{%
\begin{tabular}{lcccc}
\toprule
\textbf{System} & \textbf{Input} & \textbf{Exemplar} & \textbf{Feedback} & \textbf{Iterative} \\
\midrule
\rowcolor{lightgray}\multicolumn{5}{l}{\textit{Academic Systems}} \\
PASS~\cite{aggarwal2025pass} & Document & \ding{51}~(Script+Audio) & \ding{55} & \ding{55} \\
PresentAgent~\cite{shi2025presentagent} & Document & \ding{51}~(Video) & \ding{55} & \ding{55} \\
\midrule
\rowcolor{lightgray}\multicolumn{5}{l}{\textit{Commercial Tools}} \\
Yoodli~\cite{yoodliWebsite} & Audio/Video & \ding{55} & \ding{51}~(Numeric Metrics) & \ding{55} \\
Poised~\cite{poisedWebsite} & Audio/Webcam & \ding{55} & \ding{51}~(Dashboard Feedback) & \ding{55} \\
Orai~\cite{oraiWebsite} & Audio/Webcam & \ding{55} & \ding{51}~(Voice \& Gesture Metrics) & \ding{55} \\
\midrule
\textbf{PresentCoach (Ours)} & Slides+Audio & \ding{51}~(Slide-Aware Exemplar) & \ding{51}~(OIS Structured) & \ding{51}~(Dual-Agent Loop) \\
\bottomrule
\end{tabular}}
\label{tab:comparison}
\end{table}

\subsection{Automatic Generation of Presentation Content}

Recent work has explored the automation of presentation creation through multimodal large language models (LLMs) and vision–language systems. 
Early research on AI-assisted presentation generation primarily focused on producing concise slide decks from textual input. 
DOC2PPT~\cite{fu2022doc2ppt} and SlideGen~\cite{sefid2021slidegen} pioneered the document-to-slide generation paradigm, automatically summarizing scientific or educational texts into structured slide decks. 
Building on this foundation, later systems such as PreGenie~\cite{xu2025pregenie} and PPTAgent~\cite{zheng2025pptagent} leveraged multimodal LLMs to enhance visual design and semantic coherence, employing iterative, agent-based refinement to improve layout consistency and aesthetics.

More recent approaches move beyond static slide generation toward end-to-end multimodal presentation synthesis. 
PresentAgent~\cite{shi2025presentagent} introduced a comprehensive pipeline that transforms long-form documents into narrated presentation videos by coordinating slide composition, script generation, and audio–visual synchronization. 
Likewise, PASS~\cite{aggarwal2025pass} demonstrated automatic generation of both slides and corresponding spoken narration, realizing the concept of a “talking presentation.” 
These developments mark a clear shift from content summarization to expressive multimodal generation, yet most systems still prioritize automation and aesthetics over learning. In contrast, our approach emphasizes pedagogical value. We generates a slide-aware exemplar in the user’s cloned voice to provide a concrete, content-aligned demonstration of “what good looks like” for the user’s own slides. This exemplar establishes a clear performance target and serves as a consistent reference for subsequent practice.

\subsection{AI Coaching for Presentation}

AI-based presentation coaching has evolved from early acoustic assessment systems to multimodal, dialogue-driven tutors that actively support speaker reflection and refinement. Early research, such as ROC Speak~\cite{fung2015roc} and Presentation Trainer~\cite{schneider2015presentation}, provided prosodic and nonverbal feedback to improve pacing, tone, and posture. Later systems integrated virtual audiences and immersive environments, including Cicero~\cite{batrinca2013cicero} and the VR-based study by Chollet et al.~\cite{chollet2022training}, which compared real-time and delayed feedback. Their findings highlighted that structured, post-session feedback promotes reflection and long-term skill retention more effectively than immediate alerts.

With advances in multimodal analysis and large language models, recent approaches have shifted from quantitative scoring to qualitative, formative guidance. For instance, AI Speech Coach~\cite{wang2020voicecoach} and multimodal evaluation frameworks~\cite{wortwein2015multimodal, wortwein2015multimodal} analyze gaze, prosody, and gesture to explain how delivery affects perceived competence. Recent VR-based studies~\cite{bachmann2023virtual} further validate that simulated audiences and affective feedback can enhance user engagement and confidence. Commercial platforms such as Yoodli, Orai, and Poised~\cite{yoodliWebsite, oraiWebsite, poisedWebsite} extend these ideas to real-world applications, offering instant metrics on pacing, filler words, and confidence. However, their feedback remains largely decontextualized from slide semantics and audience perception, limiting pedagogical depth.

In contrast, our \textit{Coach Agent} reframes AI coaching as an exemplar-grounded, iterative learning process. Rather than providing isolated metrics, it analyzes user performance relative to the slide-aware exemplar and delivers multimodal, audience-aware feedback in a structured \textit{Observation--Impact--Suggestion (OIS)} format. This design bridges analytic precision with human interpretability, helping speakers understand not only \textit{what} to improve, but also \textit{why} and \textit{how}, thereby transforming presentation coaching from evaluation into guided reflection.

\section{PresentCoach System}
\subsection{System Overview}

\begin{figure*}[!t]
\centering
\includegraphics[width=\textwidth]{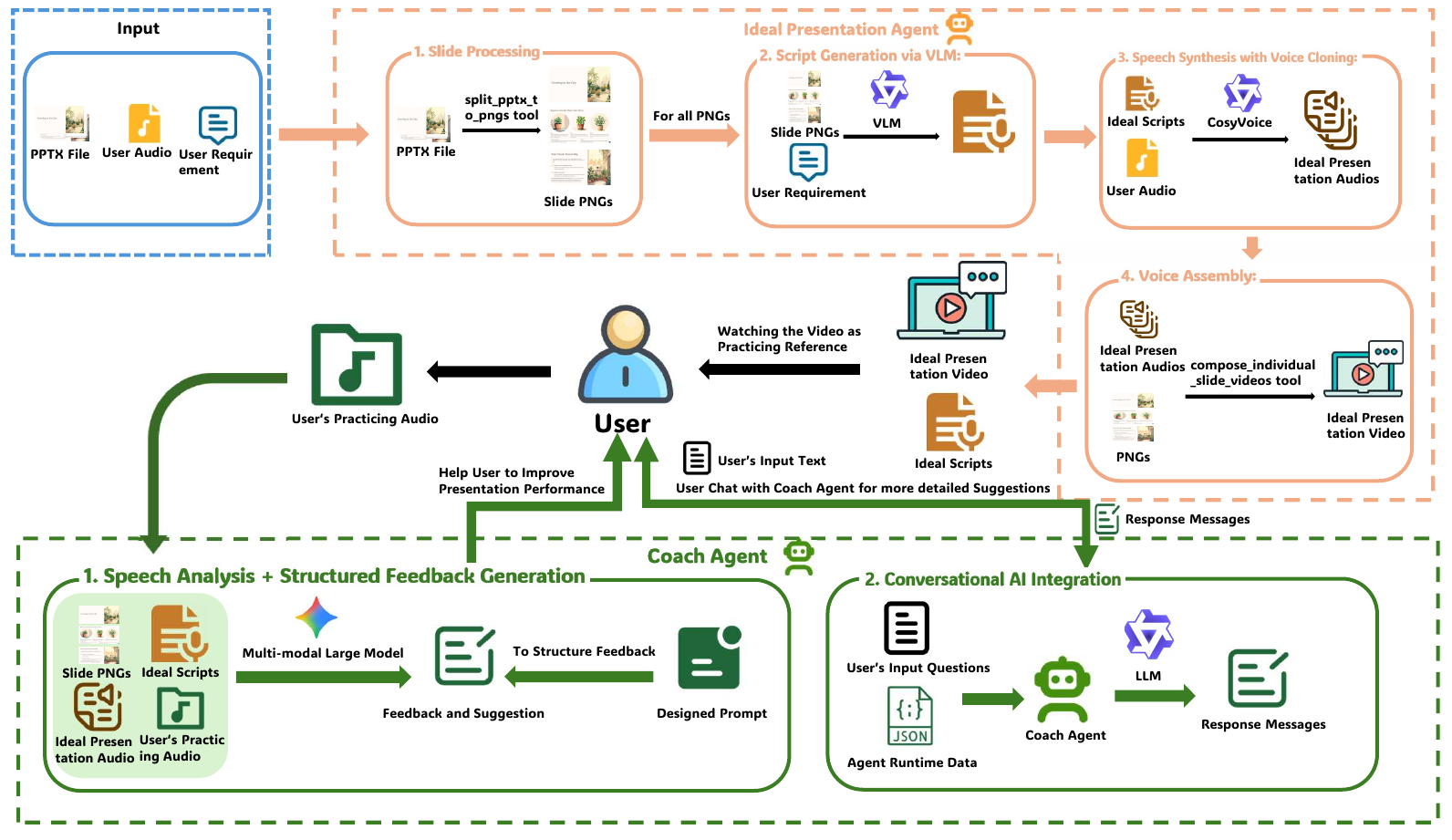}
\caption{The end-to-end workflow of the PresentCoach system. The \textit{Ideal Presentation Agent} (top) processes user inputs through a four-stage pipeline: (1) \textbf{Slide Processing} - converting .pptx to high-resolution PNGs; (2) \textbf{Script Generation} - employing a Visual Language Model (VLM) to create coherent narration scripts; (3) \textbf{Speech Synthesis} - utilizing personalized voice cloning for speech generation; and (4) \textbf{Video Assembly} - synchronizing slides and audio into a benchmark presentation. The \textit{Coach Agent} (bottom) then evaluates user practice through: (i) \textbf{Multi-modal Speech Analysis} comparing user performance against the ideal benchmark; (ii) \textbf{Structured Feedback Generation} delivering OIS-formatted suggestions; and (iii) \textbf{Conversational AI Integration} enabling interactive coaching via conversational agents.}

\label{fig:Methodology}
\end{figure*}

Our system facilitates public speaking practice through AI-powered coaching and the generation of personalized model presentations. It is composed of two primary intelligent agents: the \textit{Ideal Presentation Agent} and the \textit{Coach Agent}, as depicted in Figure~\ref{fig:Methodology}. The core of our approach is trying to offering a highly personalized and immersive learning experience that offers both example and guidance of practicing presentation. The \textit{Ideal Presentation Agent} generates a ideal presentation video delivered in the user's own cloned voice. By observing this ideal presentation video, users can more effectively identify areas for improvement. The \textit{Coach Agent} then analyzes the user's own practice recordings against this ideal model, providing targeted feedback to close the loop for accelerated skill development. The overall workflow and interaction between these agents are illustrated in Figure~\ref{fig:Methodology}.

\subsection{Ideal Presentation Agent}

The Ideal Presentation Agent is designed to automatically generate a polished, model presentation video from a standard slide deck. The input includes three parts: 1. the pptx file, 2. the user's voice, and 3. the user's requirement prompt. Its workflow is executed in four distinct stages:

\begin{enumerate}
\item \textbf{Slide Processing:} The agent begins by processing the uploaded PowerPoint files. Each slide is converted into a high-resolution PNG image using the \texttt{split\_pptx\_to\_pngs} tool. This conversion ensures visual fidelity, laying a high-quality foundation for all subsequent stages.

\item \textbf{Script Generation via Visual Language Model (VLM):} The next stage involves generating a narration script for each slide. A Visual Language Model (VLM) analyzes the generated slide images, interpreting the text, images, and layout to understand the core message. The model synthesizes a coherent English narration script (60-100 words per slide), ensuring smooth transitions between slides. The specific VLM model used here is designed to consider both the user’s prompt and a default set of instructions to create contextually accurate scripts.

\item \textbf{Speech Synthesis with Voice Cloning:} The generated narration scripts are then converted into high-quality spoken audio. This step uses a voice cloning model to replicate the user’s voice, based on a short voice sample (standardized prompt: "This is a sample text for voice cloning"). In case the voice cloning process fails or is unavailable, the system falls back on a standard text-to-speech (TTS) model to ensure continuous operation. Audio formats are automatically converted (e.g., WebM/M4A to WAV), and batch processing is used for optimized throughput.

\item \textbf{Video Assembly:} In the final stage, the agent assembles the video. The high-resolution slide images are synchronized with the corresponding synthesized audio segments using FFmpeg. A custom tool, \texttt{compose\_individual\_slide\_videos}, coordinates this process, creating a polished video that serves as the ideal model presentation for comparison and user coaching.
\end{enumerate}

\subsection{Coach Agent}

The Coach Agent is designed to help users refine their public speaking skills by providing detailed, personalized feedback on their practice sessions. It employs a comprehensive evaluation and interaction pipeline:

\begin{enumerate}
\item \textbf{Speech Analysis:} The agent evaluates a user's practice recording by conducting a multi-modal analysis. This analysis holistically assesses performance by integrating four key data sources: (1) the relevant slide image, (2) the ideal narration script, (3) the AI-generated ideal audio, and (4) the user's own recorded audio. Comparing the user's delivery against the ideal benchmark enables the generation of targeted and constructive feedback.

\noindent \textbf{Audience-Aware Feedback.} While the agent analysis the speech in a professional way, it would analysis the same speech in another way- the audience. It would simulate audience reactions by analyzing the speech audio to predict engagement and comprehension for a specific audience profile. Insights regarding content clarity and potential points of confusion are integrated into the coaching process. By synthesizing these multiple perspectives—on delivery technique from the Coach and on content reception from the Audience—the system provides the user with more comprehensive and well-rounded guidance for improvement.

\item \textbf{Structured Feedback Generation:} In order to give users more acceptable suggestions, the feedbacks would be delivered in a structured, actionable format. It begins with \textit{Sincere Encouragement}, which identifies and reinforces specific strengths in the user's performance. This is followed by \textit{OIS Feedback}, a concise (under 150 words) framework that provides a clear \textit{Observation} of an area for improvement, explains its \textit{Impact} on the presentation, and offers a concrete \textit{Suggestion} for refinement.

\item \textbf{Conversational AI Integration:} To facilitate deeper learning, the agent features a conversational coaching interface powered by Large Language Model. This interface allows users to engage in an interactive dialogue to explore their performance and feedback. By analyzing the complete session history—including previous analyses and chat logs—as input, the LLM generates highly relevant and contextual suggestions. This process enables a truly personalized coaching experience, with guidance that builds upon the user's past interactions and progress.
\end{enumerate}

\subsection{Implementation Details}

Our system's implementation is built upon a robust multi-model framework. The specific model selections for each agent are detailed below.

\subsubsection{Ideal Presentation Agent}
This agent's pipeline begins with slide analysis. We employ \textbf{Qwen2.5-VL}\cite{qwenvl25report} as the primary Visual Language Model (VLM) to interpret slide content and generate coherent narration scripts. For the critical voice cloning and synthesis stage, \textbf{CosyVoice2}\cite{cosyvoice2} serves as our core model, producing personalized speech output.

\subsubsection{Coach Agent}
The Coach Agent requires sophisticated multi-modal understanding and natural interaction. For this, we leverage \textbf{Gemini 2.5 Pro}\cite{gemini25pro} as the primary Multi-modal Large Model (MLLM) to analyze user video recordings. The interactive dialogue with the user is powered by \textbf{Qwen2.5}\cite{qwen2.5} as the primary Large Language Model (LLM).

The overall system architecture is engineered for robustness, incorporating multiple fallback mechanisms and comprehensive error handling. This design ensures reliable, high-quality output suitable for production deployment.

\section{User friendly Interface}

Our system is accessed through a sophisticated yet intuitive web-based interface, engineered according to foundational principles of Human-Computer Interaction (HCI) and user-centric design. The interface architecture is predicated on the concept of progressive disclosure, a methodology employed to mitigate cognitive load by presenting only essential information and actions at each step. This design organizes the system's complex AI-driven functionalities into a structured, three-stage workflow. This sequential paradigm is intentionally crafted to align with the user's established mental model of presentation development: from initial preparation and content generation to final practice and refinement, ensuring a seamless and comprehensible user experience.

\begin{figure*}[!t] 
  \centering
  \includegraphics[width=\textwidth]{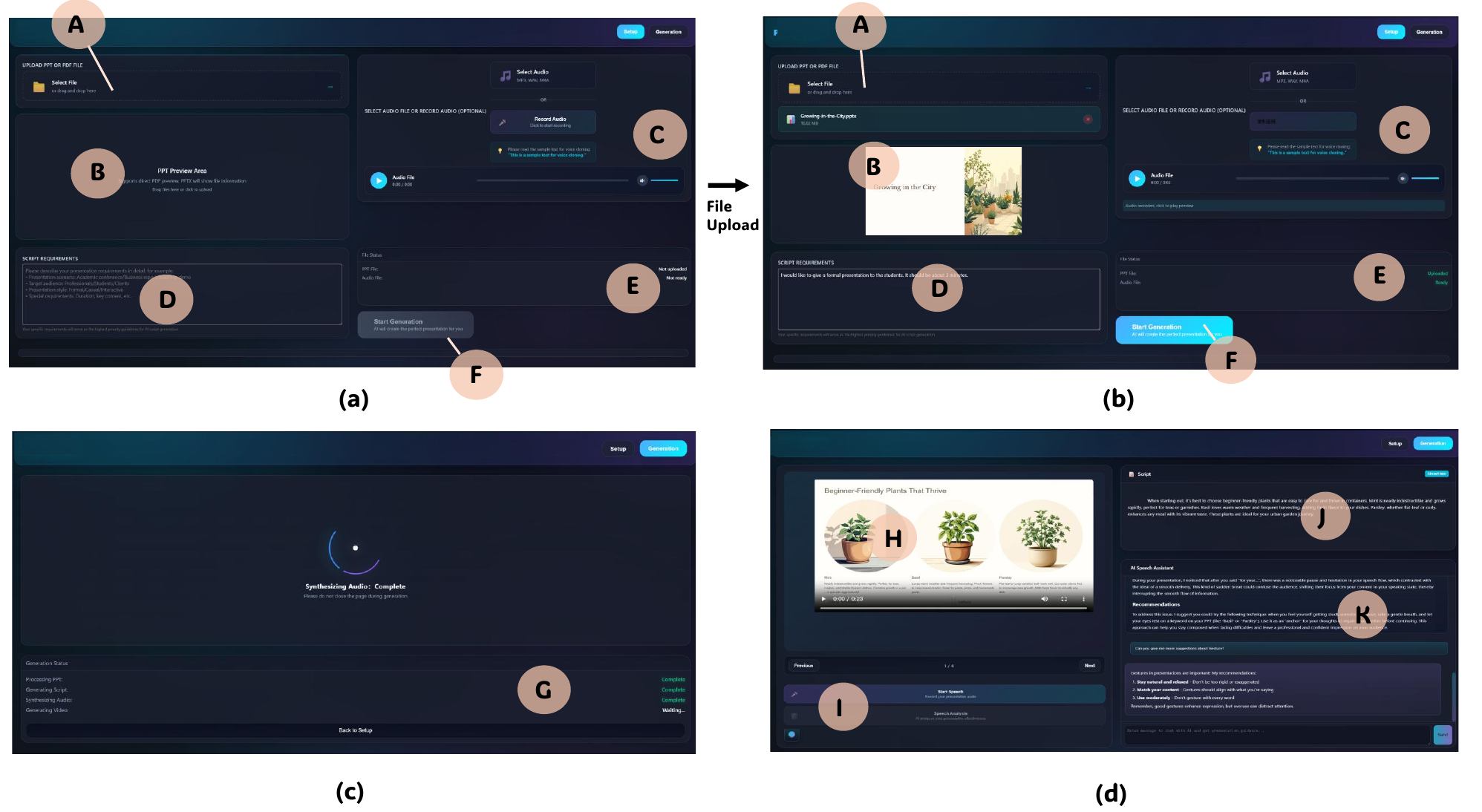}  
  \caption{(a) and (b) illustrate the Setup and Configuration stage (Stage 1), where users: A. Upload presentation files; B. Preview slide content; C. Record or upload a voice sample for cloning; D. Specify presentation requirements; E. View file upload status; F. Initiate generation. (c) visualizes the transparent pipeline of the Ideal Presentation Agent during generation (Stage 2), displaying real-time progress across key steps. (d) presents the Interactive Coaching and Practice Environment (Stage 3), which includes: H. The AI-generated ideal presentation video; I. Recording/upload controls for user practice; J. The corresponding script of the ideal narration; K. The conversational Coach Agent interface, providing structured feedback and interactive Q\&A.}
  \label{fig:ppt-talker-interface-three-stage-journey}
\end{figure*}
\subsection{Three-Stage User Journey}

The entire user experience is structured as a guided journey through three distinct stages. This scaffolded approach demystifies the AI coaching process, providing a clear, linear path from initial input to actionable feedback and skill development.

\textbf{Stage 1: Setup and Configuration} serves as the initialization phase, focusing on content ingestion and the customize the prompt. In this stage, users are prompted to upload their presentation files (e.g., PowerPoint), which form the core content for analysis. The interface then would ask users to define the presentation's user prompt, such as the intended audience (e.g., to specialist, or non-specialist) and the objective(eg. English course presentation).Another key feature of this stage is the option for voice cloning, where users can provide a short audio sample. This input is used to personalize the synthesized speech in the model presentation. The underlying design philosophy of this stage is to abstract the complexity of AI model configuration into a user-friendly, form-based interaction. By providing contextual tooltips and examples, the interface guides users on how to provide optimal inputs, thereby transforming a potentially technical setup process into an accessible and straightforward exercise.

\textbf{Stage 2: Transparent Ideal Presentation Agent Generation Process} addresses a common source of user anxiety—the opaque nature of AI computation, often referred to as the “black box” problem. Following the configuration stage, the system invokes the \textit{Ideal Presentation Agent} to automatically generate a model presentation video. During this phase, the user’s role transitions from active interaction to informed observation, while engagement is maintained through a transparent and informative visualization of system progress. Rather than presenting a generic loading bar, the interface employs a dynamic, multi-step progress indicator. Each step (e.g., “Parsing slide content,” “Generating narration script,” “Synthesizing audio track,” “Assembling video”) is explicitly labeled and visually updated upon completion. This design provides continuous system feedback that not only reassures users of correct system operation but also serves an educational function by revealing the key processes involved in generating the model presentation. Such transparency enhances user trust and deepens understanding of the AI’s workflow.

\textbf{Stage 3: Interactive Coaching and Practice Environment} represents the core learning phase and the central locus of human–AI collaboration. In this stage, users first review the AI-generated model presentation video, which is generated by the Ideal Presentation Agent in the stage 2 with integrates synchronized slide's narration. They can then record their own practice sessions within the same interface. After recording the audio of practicing presentation, the system subsequently invokes the \textit{Coaching Agent} to conduct a direct comparative analysis, enabling users to evaluate their performance against the AI-generated benchmark. The focal component of this stage is the conversational AI coach, which provides structured and personalized feedback. Through a text-based dialogue, users can explore specific feedback points, request clarification, and receive further guidance—fostering an iterative cycle of practice, reflection, and refinement. This interactive feedback mechanism transforms practice from a static activity into an adaptive and pedagogically grounded learning experience.

\subsection{Key Design Concepts}

The user interface integrates several design innovations specifically tailored to create an effective and supportive environment for presentation skill enhancement.

\textbf{User friendly Architecture} ensures that the interface remains uncluttered and task-focused. It employs a state-aware design where UI elements dynamically adapt based on the user's progress and the system's current state. For example, irrelevant controls are invisible. There are three stages in the progress, and only the corresponding tools would be shown in each stage. This design allows there would not be too much UI elements in one page, which prevents cognitive overload.

\textbf{Multi-Modal Learning Support} is a core pedagogical feature of the interface, designed to accommodate a diverse range of learning styles and preferences. This is achieved by presenting information and facilitating interaction through multiple channels:
\begin{itemize}
\item \textbf{Visual Learners} are supported through the video presentation format, side-by-side slide navigation, and visual cues within the feedback.
\item \textbf{Auditory Learners} can focus on the AI-generated narration and utilize the audio recording and playback functionalities to analyze their own vocal delivery.
\item \textbf{Interactive/Kinesthetic Learners} benefit most from the conversational, chat-based coaching module, which allows them to actively engage with and probe the AI for deeper understanding.
\item \textbf{Reading/Writing Learners} can access and analyze the generated text scripts and the written feedback provided by the coach.
\end{itemize}
This multi-modal approach enhances learning efficacy and makes the platform more inclusive and accessible.

\textbf{Intelligent Guidance} is seamlessly integrated throughout the interface via contextual tooltips, state-aware controls, and progressive feature disclosure. In alignment with HCI principles of minimizing cognitive load, information and controls are contextually presented according to the user’s current workflow stage. A three-stage progress indicator visually anchors users within the overall process, while interface elements dynamically adapt to display only task-relevant actions. Each interactive component—ranging from voice cloning to feedback review—is accompanied by concise, on-demand explanatory text, ensuring discoverability while maintaining a clean, uncluttered interface.

\subsection{Learning-Focused Design}

The interface conceptualizes presentation practice as a structured learning cycle grounded in three core pedagogical components. First, the \textbf{comparative analysis framework} enables side-by-side evaluation of user practice recordings against AI-generated exemplar presentations, rendering performance gaps both visible and measurable. Second, the \textbf{conversational coaching agent} provides personalized, OIS-formatted feedback (Observation–Impact–Suggestion) and maintains an interactive dialogue history to support longitudinal progress tracking. Third, the \textbf{multi-modal learning support} accommodates diverse learner preferences through synchronized video demonstrations, textual scripts, audio-based practice, and interactive QA modules. This integrated design fosters metacognitive engagement by guiding users to identify areas for improvement, understand the principles underlying effective presentation delivery, and monitor progress across iterative practice sessions. Rather than merely automating presentation production, the system aims to cultivate transferable presentation skills through scaffolded practice and explanatory, data-informed feedback.


\section{User Study}

We conducted a controlled user study to evaluate whether our \textit{Dual-Agent System with presentation slides (PresentCoach+PPT)} yields measurable benefits over a realistic baseline that relies solely on slides (\textit{PPT-only}). 
The study examined both affective and experiential outcomes under conditions closely mirroring authentic short-presentation rehearsal. 
By situating the evaluation in a realistic but controlled environment, we aimed to assess how intelligent dual-agent coaching can enhance presentation preparation compared with self-directed practice.

\subsection{Participants}

We aimed to evaluate how non-native English speakers engage with our \textit{Dual-Agent Rehearsal System with slides (PresentCoach+PPT)}. 
We recruited 24 participants (10 female, 14 male; age range: 19–25, $M_{age}=22.38$, $SD=1.97$) from the same university. 
All were native Mandarin speakers who used English as a second language (L2) for academic or professional presentations. 
Participants’ self-reported IELTS speaking scores ranged from 6.0 to 6.5 ($M=6.29$, $SD=0.25$), indicating competent but non-fluent proficiency. 
This homogeneous linguistic background ensured comparability while representing the target population of \textit{PresentCoach} users. 
Detailed demographic information for all participants is provided in the Appendix~\ref{appendix:demographics} (Table~\ref{tab:participant_demographics}).

\subsection{Task Design}

The study employed a standardized English presentation rehearsal task designed to induce a moderate yet realistic level of public-speaking anxiety. This design allowed us to examine whether \textit{PresentCoach} could effectively reduce presentation-related anxiety compared with self-directed rehearsal. The task simulated a typical short-presentation scenario that university students and early-career professionals often encounter. All aspects of materials, timing, and evaluation were standardized to ensure that any observed effects were primarily due to the rehearsal condition.

\paragraph{Materials.}
A pool of eight neutral, informational English topics was prepared, each accompanied by a three-slide PowerPoint deck and concise bullet-point notes outlining key ideas (e.g., “The Pyramid Principle of Healthy Eating” and “Five Strategies for Effective Time Management”). For each participant, two distinct topics were randomly selected: one for the \textbf{Pre-test} and another for the \textbf{Post-test}. Using different topics helped to avoid learning effects that commonly occur in repeated-measures designs, ensuring that changes in anxiety reflected the rehearsal condition rather than topic familiarity. Participants were instructed not to use scripts, teleprompters, or online resources, maintaining the authenticity of their delivery. Each presentation lasted three minutes, long enough to engage attention and self-awareness but short enough to avoid fatigue or memorization.

\paragraph{Task Objective.}
The primary goal of the task was to measure whether rehearsal with PresentCoach could reduce self-perceived anxiety during speech delivery. The three-minute format was chosen to balance realism and controllability, allowing for clear measurement of changes in confidence and composure between pre-test and post-test sessions.

\subsection{Procedure}

Each experimental session lasted approximately 70 to 75 minutes and followed a consistent three-phase sequence. At the beginning, participants provided informed consent, completed a short demographic questionnaire, and received an explanation of the study procedure.

\textbf{(1) Pre-test:} Participants reviewed their assigned slides for two minutes and then delivered a three-minute English presentation. This performance served as the baseline measurement of self-perceived anxiety.  

\textbf{(2) Rehearsal Phase:} Participants rehearsed according to their assigned condition. Those in \textbf{Group A (PresentCoach+PPT)} received a five-minute introduction and practiced for twenty minutes with PresentCoach, which provided dual-agent modeling and feedback designed to encourage reflection and confidence building. Those in \textbf{Group B (PPT-only)} rehearsed independently for the same duration with identical slides and notes but without any AI assistance. The post-test topic was predetermined to ensure comparability and to avoid repetition across sessions.

\textbf{(3) Post-test and Debriefing:} Immediately after the rehearsal, participants delivered a new three-minute presentation using the assigned post-test slides. This second performance captured the post-rehearsal state and allowed for a direct comparison with baseline measures. Following the post-test, participants completed the questionnaires and took part in a short semi-structured interview reflecting on their rehearsal experiences, changes in confidence or anxiety, and perceptions of usability and engagement.

All sessions were conducted in a quiet laboratory setting, using identical recording equipment and seating arrangements. Participants were reminded not to access external aids during the task. The procedure ensured both ecological realism and experimental control, supporting a fair comparison between AI-assisted and self-directed rehearsal.

\subsection{Measurement}

\paragraph{\textbf{RQ1: Reduction of Public-Speaking Anxiety.}} 
Speaking confidence and anxiety were measured using the \textit{Personal Report of Confidence as a Speaker (PRCS)} \cite{paul1966insight}, administered before and after the experiment to capture pre–post changes. This measure provided the primary quantitative indicator for evaluating whether PresentCoach effectively reduced self-reported anxiety during presentation rehearsal.

\paragraph{\textbf{RQ2: Cognitive Workload and System Usability.}} 
After completing the post-test, participants completed two standardized instruments. The \textit{NASA Task Load Index (NASA-TLX)} \cite{hart1988development} assessed perceived workload across six dimensions: mental demand, physical demand, temporal demand, performance, effort, and frustration. Each item was rated on a 7-point Likert scale. In addition, a five-item subset of the \textit{System Usability Scale (SUS)} \cite{brooke1996sus} evaluated usability and overall interaction fluency on a 5-point Likert scale. Together, these measures assessed whether PresentCoach maintained a cognitively manageable workload and satisfactory usability during practice.

\paragraph{\textbf{RQ3: Perceived Dual-Agent Interaction.}} 
To explore how users perceived the collaboration between the two agents, we designed a custom eight-item, 7-point Likert questionnaire. The items assessed perceived role complementarity, motivational influence, and feedback clarity. Participants also took part in short semi-structured interviews, which provided qualitative insights into their interaction experience, emotional responses, and learning processes. These qualitative findings complemented the quantitative data by clarifying how users interpreted and experienced the distinct functions of the two agents.

\paragraph{Statistical Analysis.} 
All quantitative analyses were conducted to evaluate changes in user public-speaking anxiety and to summarize perceived workload, usability, and interaction experience. For PRCS scores, distributional assumptions were examined using the \textit{Shapiro–Wilk test} \cite{shapiro1965analysis}. When normality was not supported, the \textit{Wilcoxon signed-rank test} \cite{wilcoxon1945individual} was used to assess pre–post differences within each group. For the NASA-TLX and SUS measures, descriptive statistics (means and standard deviations) were reported to illustrate that the system maintained acceptable levels of cognitive demand and usability. Responses from the dual-agent experience questionnaire were analyzed descriptively to highlight central tendencies and patterns. No multiple-comparison correction was applied, as inferential testing was limited to the PRCS measure.

\section{Findings}
\label{sec:findings}

This section presents the main findings of our study, organized according to the three research questions. Quantitative analyzes address the measured changes in speaking confidence, perceived workload, and usability, followed by qualitative insights that explain participants’ perceptions of the dual-agent interaction. All results are reported as mean $\pm$ SD, and inferential statistics were applied only to the PRCS data.

\subsection{Quantitative Results}

\subsubsection{\textbf{RQ1: Reduction of Public-Speaking Anxiety}}

We first examined whether rehearsal with \textit{PresentCoach+PPT} reduced public-speaking anxiety compared with \textit{PPT-only} self-practice. Participants’ PRCS scores showed clear improvement following dual-agent rehearsal. In the PPT-only group. Wilcoxon signed-rank tests confirmed that the pre--post improvement was statistically significant in the PresentCoach+PPT group ($p = .016$) but not in the PPT-only group ($p = .219$). Between-group comparison at post-test further revealed that the PresentCoach+PPT group reported significantly higher confidence than the PPT-only group ($t(22) = 2.83$, $p = 0.014$). Table 2 summarizes the full results.

These results suggest that PresentCoach’s feedback and guided rehearsal contributed to a measurable reduction in anxiety and higher self-reported confidence. Several participants also noted in written comments that the session felt “less intimidating” and “more focused,” attributing their confidence gains to the system’s structured, constructive feedback process.

\begin{table*}[ht]
\centering
\caption{Within- and between-group comparisons of PRCS scores (5-point Likert). Wilcoxon signed-rank tests assess pre–post changes within each group; independent-sample $t$-tests compare groups at post-test.}
\label{tab:prcs_full}
\begin{tabular}{llcccccc}
\toprule
\textbf{Measure} & \textbf{Group} & \textbf{Pre-test (M±SD)} & \textbf{Post-test (M±SD)} & \textbf{$\Delta$ (\%)} & \textbf{$p_{\text{within}}$} & \textbf{$t_{\text{post}}$} & \textbf{$p_{\text{post}}$} \\
\midrule
\multirow{2}{*}{PRCS} 
 & PresentCoach+PPT & 3.69 ± 0.81 & 5.03 ± 0.63 & +36.3\% & 0.016 & \multirow{2}{*}{2.83} & \multirow{2}{*}{0.014} \\
 & PPT-Only & 3.18 ± 0.65 & 3.69 ± 1.17 & +16.1\% & 0.219  &  &  \\
\bottomrule
\end{tabular}

\vspace{2mm}
\parbox{0.9\textwidth}{
\textit{Note.} 
$\Delta$ (\%) represents the percentage change in the group-level mean PRCS score from pre-test to post-test, calculated as 
$\Delta = \frac{(\textit{Mean}_{\text{Post}} - \textit{Mean}_{\text{Pre}})}{\textit{Mean}_{\text{Pre}}} \times 100$. 
A positive $\Delta$ indicates an increase in mean self-confidence and a corresponding reduction in speaking anxiety. 
$p_{\text{within}}$ values were obtained using Wilcoxon signed-rank tests comparing pre- and post-test scores within each group. 
$t_{\text{post}}$ and $p_{\text{post}}$ denote independent-sample $t$-test statistics comparing the two groups at post-test.
}
\end{table*}

\subsubsection{\textbf{RQ2: Cognitive Workload and System Usability}}
\begin{figure*}[t]
  \centering
  \includegraphics[width=0.9\linewidth]{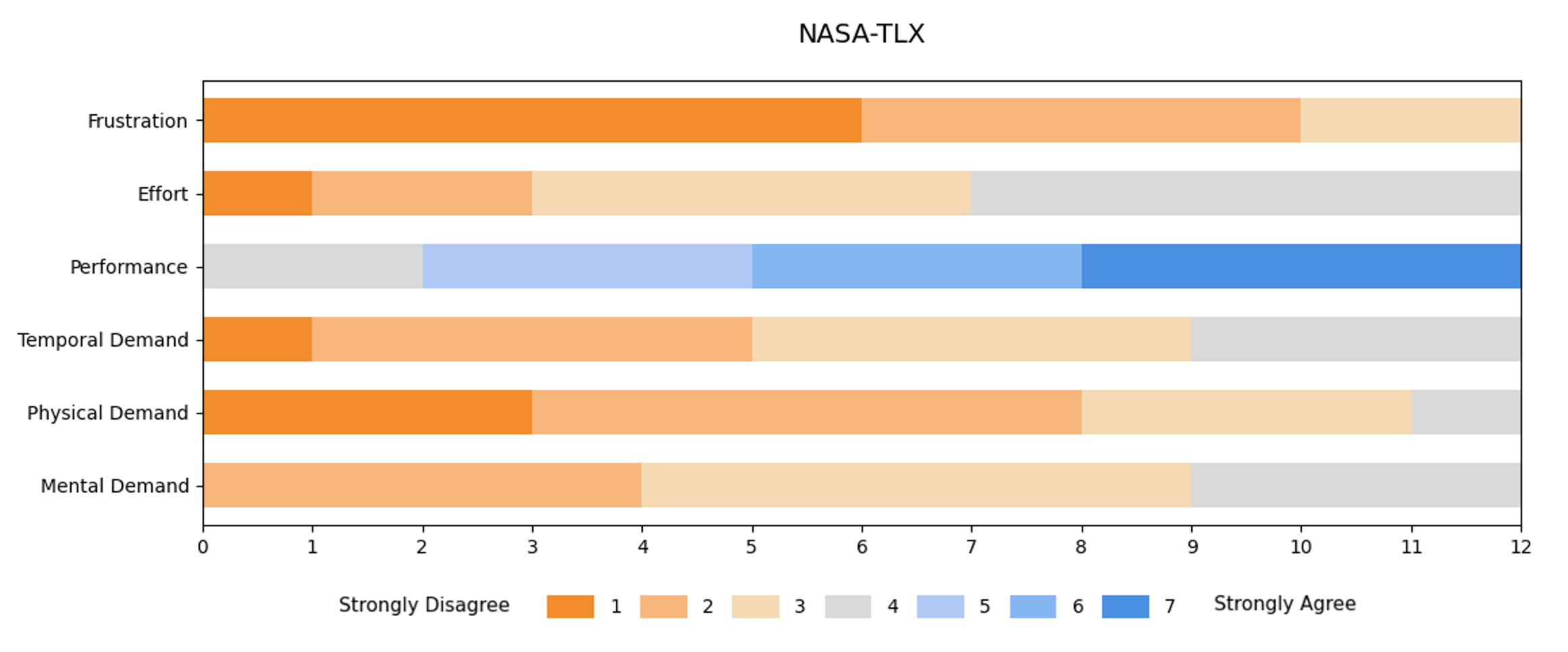}
  \caption{Perceived cognitive workload across six NASA-TLX dimensions (Mental, Physical, and Temporal Demand, Performance, Effort, and Frustration). Ratings were given on a 7-point scale, showing that most participants experienced the task as cognitively engaging but not overwhelming.}
  \label{fig:nasa_tlx_results}
\end{figure*}

We next analyzed whether the feedback-rich interaction in PresentCoach maintained an acceptable cognitive workload and provided a satisfactory usability experience. Post-test ratings on the NASA-TLX (Figure~\ref{fig:nasa_tlx_results}) indicated moderate overall workload ($M=2.9$, $SD=0.8$ on a 7-point scale), below levels typically associated with cognitive overload. Mental Demand and Effort received the highest subscale scores, reflecting the active engagement required to process and apply feedback, while Physical Demand and Frustration remained low. These results suggest that participants experienced PresentCoach as cognitively demanding but manageable, consistent with a productive learning effort.

\begin{figure*}[t]
  \centering
  \includegraphics[width=0.9\linewidth]{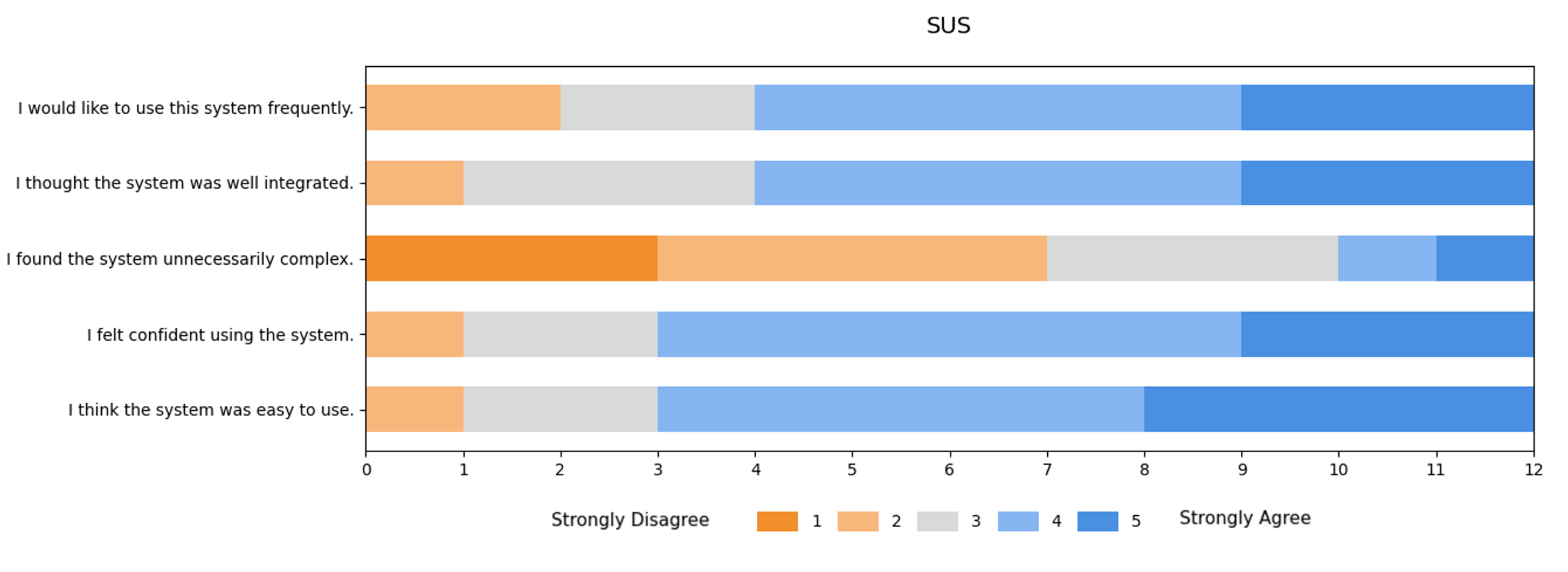}
    \caption{Participants’ usability ratings for the PresentCoach system on a 5-point Likert scale. Each item reflects an aspect of perceived ease of use, integration, and confidence. Overall responses indicate that participants found the system intuitive and well integrated after brief familiarization.}
    
  \label{fig:sus_results}
\end{figure*}

Usability scores on the five-item SUS subset (Figure~\ref{fig:sus_results}) were high ($M=4.1$, $SD=0.5$ on a 5-point scale), exceeding conventional benchmarks for satisfactory usability. Participants described the interface as coherent and the conversational flow as natural, reporting that the transitions between the two agents became intuitive after minimal familiarization. Open-ended remarks highlighted the clarity of the Coach Agent’s feedback and the motivating tone of the Ideal Presentation Agent as the main contributors to ease of use. Overall, these results show that PresentCoach successfully balanced feedback complexity with interface simplicity.

\subsubsection{\textbf{RQ3: Perceived Dual-Agent Experience}}
\begin{figure*}[t]
  \centering
  \includegraphics[width=0.9\linewidth]{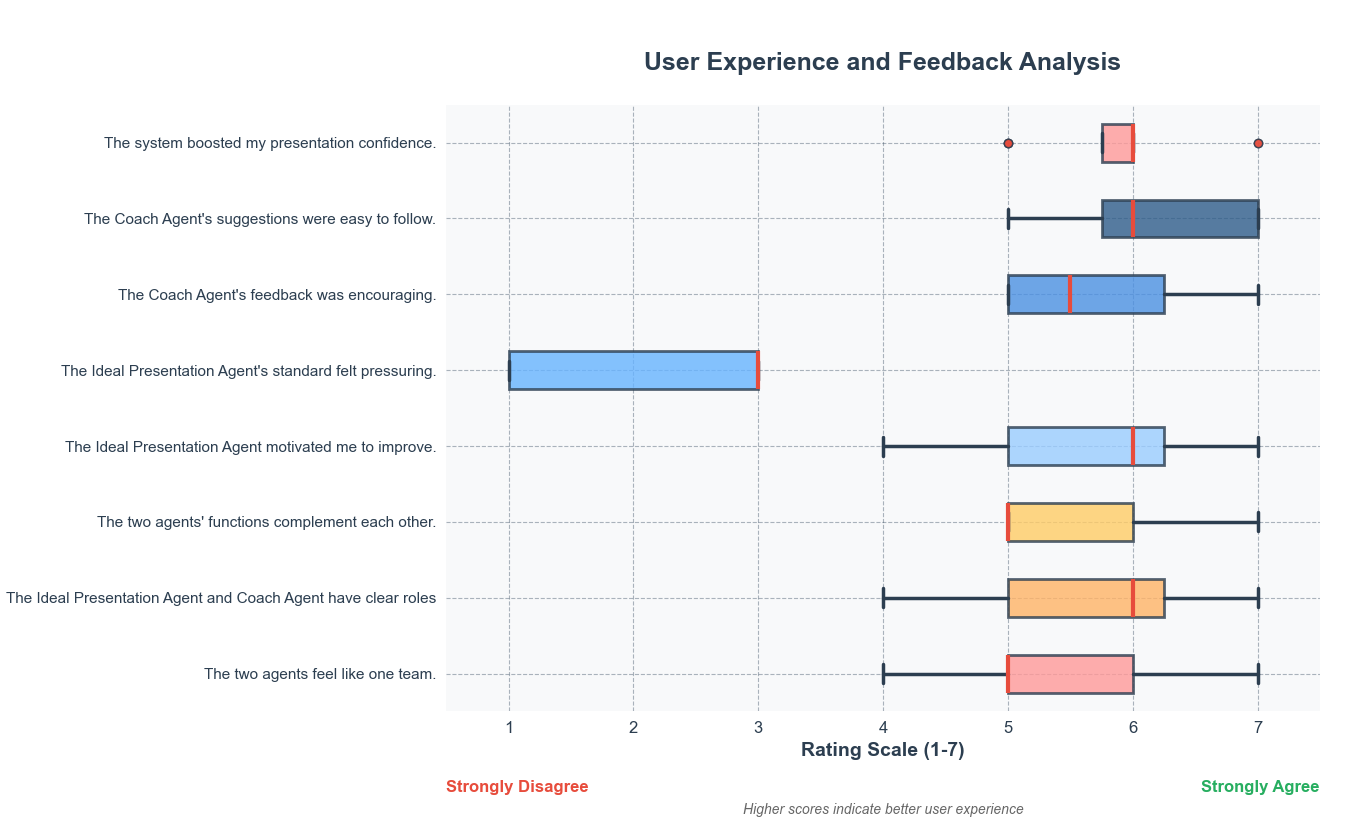}
  \caption{Participants’ evaluations of the dual-agent interaction on a 7-point Likert scale. Responses show that users clearly distinguished the roles of the Ideal Presentation Agent and the Coach Agent, perceived their collaboration as complementary, and regarded the overall experience as motivating and supportive.}
  \label{fig:user_feedback}
\end{figure*}

To further examine how participants perceived the interaction between the two agents, we analyzed responses to the seven-item, 7-point Likert questionnaire on dual-agent collaboration (Figure~\ref{fig:user_feedback}). Participants clearly distinguished the roles of the Ideal Presentation Agent and the Coach Agent while perceiving them as complementary. Items such as “I can clearly distinguish the two agents’ roles” ($M=6.3$, $SD=0.6$) and “The two agents complemented each other” ($M=6.1$, $SD=0.7$) received the highest ratings. Feedback-related statements also scored positively: “The Coach Agent’s feedback was clear and specific” ($M=5.8$, $SD=0.8$) and “The feedback helped me identify how to improve my presentation” ($M=6.0$, $SD=0.6$). Only a few participants reported mild pressure when comparing themselves to the Ideal Agent ($M=3.4$, $SD=1.1$), although most interpreted this as motivational rather than discouraging. These findings suggest that participants not only understood the functional distinction between the two agents but also viewed their interplay as a coherent, supportive process for self-improvement.

\subsection{Qualitative Results}

To deepen our understanding of how participants experienced the dual-agent rehearsal, we analyzed semi-structured interviews conducted after the post-test. Each session lasted about 10 to 15 minutes and explored how participants interpreted the agents’ roles, how they used feedback during rehearsal, and how these experiences shaped their confidence and engagement. The interviews were conducted in Chinese and later translated into English. Two bilingual authors independently translated and cross-checked the transcripts to ensure the accuracy of quoted expressions. Transcripts were coded inductively to identify recurring patterns.

\subsubsection{\textbf{Perception of the Ideal Presentation Agent}}
Participants generally described the Ideal Presentation Agent as a motivating reference rather than a source of pressure. Seeing an ``ideal version" of themselves perform confidently helped establish a tangible target and reinforced their belief that improvement was achievable. One participant noted, \textit{``English isn’t my first language, so I often hesitate or stumble when speaking. But hearing my own voice deliver a speech fluently, with a clear accent and emotional tone, was fascinating...it made me feel that kind of expressiveness was truly within reach" (P02)}. Another shared, \textit{``You reproduced my voice really well. When I first heard it, I was a bit surprised...the English and pronunciation sounded so natural that I almost thought it was a real speech I had given before." (P07).} Several participants (N=8) mentioned revisiting short segments of the video before practice to recall phrasing or pacing patterns, and even to imitate the tone and emotional expression in the voice, suggesting that the model provided both cognitive and emotional guidance.

\subsubsection{\textbf{Perception of the Coach Agent}}
The Coach Agent was consistently perceived as a constructive, empathetic listener who offered clear and actionable feedback. Participants valued its specific advice on timing, pauses, and filler-word reduction, often describing it as ``accurate but not harsh." One participant explained, \textit{``Its feedback was professional and kind. I never ask my classmates to rehearse with me or give feedback on my speech because I’m afraid they might laugh at my English pronunciation...but with the AI, I didn’t feel that way." (P04).} Another remarked, \textit{``I felt it was like both a teacher and an audience. When the agent started listening, I wanted to take the practice more seriously, as if someone was really there. Before, when I practiced alone, I just memorized my script and didn’t think much about pauses or emotions, but with the agent, each practice felt like a rehearsal." (P09).} The system’s supportive tone encouraged self-reflection and experimentation rather than defensiveness, enabling participants to internalize feedback as useful guidance.

\subsubsection{\textbf{Synergy Between the Two Agents}}
A recurring theme across interviews was the perceived complementarity between the two agents. Participants explained that the Ideal Agent provided a concrete model of effective delivery, while the Coach Agent contextualized that model by identifying personal gaps. Many (N=10) said that feedback became clearer after recalling how the Ideal Agent handled similar phrases or moments. As one participant summarized, \textit{``If there were only the ideal agent, my practice might still lack professional feedback. On the other hand, if there were only the coach agent, I might not have a clear reference to follow. It could give advice, but I still wouldn’t know exactly how to practice. So I think both agents are really important.”} (P03). Another commented, ``It’s like seeing what good looks like, then learning how to reach it" (P05). This dynamic interaction between demonstration and feedback was widely recognized as a defining strength of the system.

\subsubsection{\textbf{Summary}}
Overall, participants viewed the dual-agent design as a balanced and supportive rehearsal structure: one agent inspired motivation through ideal modeling, and the other offered empathetic, targeted feedback. This combination fostered emotional safety and sustained engagement, helping participants approach practice with confidence and curiosity. The qualitative findings thus help explain the anxiety reduction and positive user experience observed in the quantitative results.

\section{Discussion}
This discussion section synthesizes the implications of our findings. First, we elucidate the psychological mechanisms through which our system enhances learner motivation and self-confidence, positioning its contribution relative to prior work. Subsequently, we reflect on the limitations of the current study and the broader pedagogical considerations, which also serve to outline promising avenues for future research.

\subsection{The Motivational Role of the Ideal Agent}
Our findings demonstrate that the Ideal Agent substantially enhances learners’ presentation motivation and self-confidence through two core psychological mechanisms. In contrast to prior systems which offer only isolated metrics, our solution moves beyond this limitation by analyzing the user's presentation performance in the context of the slides and delivering dynamic, multimodal feedback. First, hearing one's own cloned voice creates a stronger sense of self-identification and psychological proximity compared to external exemplars, making the target performance feel personally attainable rather than abstract. Second, the agent transforms vague presentation goals into concrete, visualized models, reducing the cognitive load associated with skill acquisition and increasing self-efficacy. This aligns with recent voice-based learning research\cite{wang2020voicecoach} demonstrating the motivational advantages of self-referential auditory stimuli. The voice-based implementation establishes a foundation for multi-modal extension—future iterations could generate ideal avatars modeling body language and eye contact, providing learners with a comprehensive "ideal self" to emulate across communication channels.

\subsection{Limitation}
While our dual-agent system demonstrates promising results in presentation skill development, several limitations merit consideration. 


\textbf{Evaluation Scope:} Our user study focused on short-form presentations (3 minutes) with standardized slide decks among L2 English speakers with specific proficiency profiles. The generalizability of our findings to longer, more complex presentations, native speakers, or different language pairs remains to be established. Furthermore, the study duration was insufficient to assess long-term skill retention and transfer to real-world presentation contexts.

\textbf{Pedagogical Considerations:} While the feedback provided in this study offers valuable, data-driven guidance, it is important to note that our current analysis is based solely on the user's speech. This means that crucial elements of live presentations, such as body language, eye contact, and audience engagement, fall outside the scope of the current system's analysis. Integrating these multimodal cues constitutes a vital direction for our future research.

\subsection{Future Work}
\textbf{Future Directions:}

Our findings indicate that the Coach agent not only provides instructional feedback but also functions as a responsive audience, thereby enriching the practice experience. Building on this dual role, we plan to enhance the agent’s realism by introducing interactive virtual audiences capable of reactive behaviors during presentations, such as responsive nodding and eye contact tracking. These features are designed to foster a more empathetic and engaged virtual environment, with the specific goal of improving users’ visual engagement and boosting their self-confidence in public speaking. What's more, future iterations could generate idealized avatars that model appropriate body language and eye contact, providing learners with a comprehensive "ideal self" to emulate across multiple communication channels.

We also plan to personalize feedback even further by adapting the agent’s responses to each user’s individual progress and challenges. This would allow for more tailored goal-setting and guidance. Another exciting direction involves preparing users for a variety of real-world communication scenarios by introducing cross-cultural presentation styles. Long-term, we want to track users' progress over time, providing insights into their growth and skill development.

Lastly, we’re focused on improving the system's efficiency through techniques like model distillation, which will enhance performance and accessibility, especially as we scale the system to support more learners.

\section{Conclusion}

This paper introduced a dual-agent system that addresses a key limitation in automated presentation training: the disconnect between abstract evaluation and concrete improvement. Our core contribution lies in synergizing an \textit{Ideal Presentation Agent}, which generates a slide-aware, personalized presentation example video using the learner's own voice, with an interactive \textit{Coach Agent} that provides context-aware feedback. This approach moves beyond isolated metrics by offering a tangible target performance and explanatory guidance, which our study confirms significantly enhances learners' delivery and self-confidence. By successfully bridging demonstration with dialogue, our work establishes a new paradigm for AI systems that not only assess but also holistically coach complex presentation skills.

\bibliographystyle{ACM-Reference-Format}
\bibliography{reference}

\end{document}